\documentclass[aip, preprint, longbibliography]{revtex4-1}

\usepackage{graphicx}
\usepackage{subfigure}
\usepackage{ulem}
\usepackage{color}
\usepackage{bm}
\usepackage{comment}
\usepackage{soul}
\usepackage{soul,xcolor}

\usepackage{todonotes}
\definecolor{electricpurple}{rgb}{0.75, 0.0, 1.0}

\usepackage{amsmath}

\begin{document}

\setstcolor{electricpurple}



\title{Valley based splitting of topologically protected helical waves in elastic plates}

\author{M. Miniaci$^{*}$}
\affiliation{School of Aerospace Engineering, Georgia Institute of Technology, GA 30332 Atlanta, USA}
\affiliation{EMPA, Laboratory of Acoustics and Noise Control, \"Uberlandstrasse 129, 8600 D\"ubendorf, Switzerland}
\author{R. K. Pal}%
\affiliation{School of Aerospace Engineering, Georgia Institute of Technology, GA 30332 Atlanta, USA}
\author{R. Manna}
\affiliation{School of Aerospace Engineering, Georgia Institute of Technology, GA 30332 Atlanta, USA}
\author{M. Ruzzene}%
\affiliation{School of Aerospace Engineering, Georgia Institute of Technology, GA 30332 Atlanta, USA}
\affiliation{School of Mechanical Engineering, Georgia Institute of Technology, GA 30332 Atlanta, USA}
\email{marco.miniaci@gmail.com, ruzzene@gatech.edu}

\date{\today}


\maketitle 


\textbf{Topological protection~\cite{hasan2010colloquium} offers unprecedented opportunities for wave manipulation and energy transport in various fields of physics, including elasticity~\cite{susstrunk2015observation, mousavi2015topologically, MiniaciPRX2018, susstrunk2016classification}, acoustics~\cite{yang2015topological, Wangeaaq1475}, quantum mechanics~\cite{Chiu2016Classification} and electromagnetism~\cite{khanikaev2013photonic, lu2014topological, OzawaArxiv2018}.
Distinct classes of topological waves have been investigated by establishing analogues with the quantum~\cite{nash2015topological}, spin~\cite{He_NatPhys_2016} and valley Hall~\cite{lu2017observation} effects. We here propose and experimentally demonstrate the possibility of supporting multiple classes of topological modes within a single platform.
Starting from a patterned elastic plate featuring a double Dirac cone, we create distinct topological interfaces by lifting such degeneracy through selective breaking of symmetries across the thickness and in the plane of the plate. We observe the propagation of a new class of heterogeneous helical-valley edge waves capable of isolating modes on the basis of their distinct polarization. 
Our results show the onset of wave splitting resulting from the interaction of multiple topological equal-frequency wave modes, which may have significance in applications involving elastic beam-splitters, switches, and filters.}

Interfaces between distinct topological phases of matter support exotic localized wave modes that allow defect-immune, lossless energy transport~\cite{mousavi2015topologically, MiniaciPRX2018}.
Distinct classes of topological phases exist depending on the dimension and the symmetries associated with different interface modes\cite{OzawaArxiv2018}.
Examples in two dimensions include analogues of the quantum Hall, spin Hall and valley Hall effects, supporting chiral, helical and valley modes, respectively~\cite{fleury2014sound, souslov2017topological, huber2016topological, ningyuan2015time, pal2017edge, pal2016helical, brendel2017pseudomagnetic, Li2017Weyl, zhu2018design, MiniaciPRX2018}.
While chiral modes require the breaking of time-reversal symmetry, helical and valley modes involve solely passive components and arise from the breaking of geometrical symmetries in lattices whose reciprocal space is characterized by singularities such as double Dirac cones\cite{sakoda2012double} and Weyl points\cite{Lu622Science2015}. Recent studies have indicated that novel physical phenomena may arise from the interaction of distinct classes of topological modes~\cite{Nieaap8802, LaiArxiv2017}. Indeed, while structures supporting chiral, helical and valley modes separately have been broadly investigated, the implementation of a single platform supporting multiple classes of such modes has not been illustrated yet. This is a particularly challenging task for mechanical substrates as in the case of elastic plates, due to the presence of multiple guided wave modes and their tendency to hybridize at interfaces and free boundaries~\cite{graff2012wave}. We here report on an elastic plate capable of hosting purely helical and heterogeneous helical-valley modes. Numerical models and experimental implementations investigate the interaction of helical edge waves at interfaces between configurations that are topologically distinct. Through this platform, we demonstrate the ability to split equal-frequency helical edge waves differing on the basis of their polarization when they impinge on distinct interfaces at a common junction. 

We consider an elastic plate patterned with a periodic array of through-the-thickness circular and triangular holes~\cite{MiniaciPRX2018}, as shown in Fig. \ref{fig1}a. The plate is periodic along the directions defined by the $\mathbf{a}_1$ and $\mathbf{a}_2$ vectors. Its band structure exhibits an isolated double Dirac cone at the $K$ point as illustrated by the dispersion curves denoted by the red circles in Fig.~\ref{fig1}d. The Dirac cones arise as a result of the $D_{3h}$ symmetry of the structure, i.e. consisting of $C_3$ (three fold rotational) symmetry, $\sigma_h$ symmetry (or reflection symmetry about the mid-plane of the plate), and $\sigma_v$ symmetry (or inversion symmetry about a plane normal to the mid-plane of the plate and along the lattice vectors).

Starting from this configuration, geometric perturbations are introduced so to break the $\sigma_h$ and $\sigma_v$ symmetries, and produce nontrivial bandgaps that respectively support helical and valley modes in a common frequency range. Specifically, we break the $\sigma_h$ symmetry by replacing the through holes with blind holes of height $h$, as shown in Fig. \ref{fig1}b. We denote the configuration with the blind holes on the top (bottom) surface as $H^+$ ($H^-$). This geometric perturbation causes modes spanning the two Dirac cones to hybridize in analogy with the spin-orbital coupling interaction in QSHE, which breaks the degeneracy and opens a topological bandgap (Fig.~\ref{fig1}e). The interface between $H^+$ and $H^-$, here denoted as $\mathcal{I}(H^+,H^-)$, separates phases that are inverted ($\sigma_h$-transformed) copies of each other, and supports two helical edge modes spanning the gap with positive ($\Phi^+$) and negative ($\Phi^-$) group velocity, respectively (Fig. \ref{fig2}a).

Next, we break the $\sigma_v$ while preserving $C_3$ and $\sigma_h$ symmetries, by considering holes in each unit cell of different radii, namely $r$ and $R$. This leads to two distinct phases, denoted as $V^r$ and $V^R$ (Fig.~\ref{fig1}c). Contrary to the previous case, an interface that separates two $\sigma_v$-transformed copies of the structure supports a single valley mode, with positive or negative group velocity, depending on the type of interface, namely $\mathcal{I}(V^r,V^r)$ or $\mathcal{I}(V^R,V^R)$ with two adjacent holes of diameter $r$ or $R$, respectively. The existence of these edge modes is a consequence of the bulk-boundary correspondence principle~\cite{LaiArxiv2017} and can be predicted by computing the valley Chern numbers. Although the total Chern number is zero in each band, the Chern number computed around the $K$ and $K'$ points will have non-zero values\cite{vila2017observation}. Based on these assumptions, we can infer that an interface between structures supporting helical and valley modes will still support a single hybrid edge mode, named helical-valley (HV) mode hereafter, with either positive ($\Psi^+$), as in the case of $\mathcal{I}(H^+,V^R)$ (Fig.~\ref{fig2}c), or negative ($\Psi^-$) group velocity for $\mathcal{I}(H^-,V^r)$ (Fig.~ \ref{fig2}e).

The existence of the above mentioned hybrid HV edge modes is verified through the computation of the band structure of finite strips including a total of $20 \times 1$ unit cells, with periodicity conditions imposed along the $\mathbf{a}_1$ direction and free boundaries along $\mathbf{a}_2$ (see Methods for details on computations). Results are reported in Figs. \ref{fig2}a, \ref{fig2}c and \ref{fig2}e, where the bulk modes are shaded in gray, while the edge states are denoted by the black, blue and red circles for the $\mathcal{I}(H^+,H^-)$, $\mathcal{I}(H^+,V^R)$ and $\mathcal{I}(H^-,V^r)$, respectively. The additional notation of the indexes $+ / -$ for the modes keeps track of their different group velocity with respect to the direction of propagation. As noted above, two edge modes $\Phi^{+,-}$ are supported by the $\mathcal{I}(H^+,H^-)$ interface, while a single mode exist at the $HV$ interfaces: $\Psi^+$ and $\Psi^-$ for $\mathcal{I}(H^+,V^R)$ and $\mathcal{I}(H^-,V^r)$, respectively.

A domain wall formed according to each of the three interfaces considered, i.e. $\mathcal{I}(H^+,H^-)$, $\mathcal{I}(H^+,V^R)$ and $\mathcal{I}(H^-,V^r)$, separates two phases in the middle of the strip. Let us consider the two edge waves that initially propagate along the $\mathcal{I}(H^+,H^-)$ interface and subsequently encounter two $\mathcal{I}(H,V)$ interfaces, each supporting a single HV mode with distinct polarization. At the $y$-junction, each wave follows the interface that matches its polarization, thus causing the two wave modes to split. This is possible under the condition that the frequencies for the 3 edge states match. To ensure this, the bandgaps of the $H$ and $V$ lattices are designed to occur in a common range of frequencies highlighted by the gray rectangles in Figs. \ref{fig1}e,f which is achieved by properly selecting the geometric perturbations that produce the distinct topological phases. Specifically, the results for the $H$ phase (Fig. \ref{fig1}e)  correspond to a blind hole depth $h=0.91 H$, with $H$ denoting the plate thickness, while the $V$ phase results (Fig. \ref{fig1}f) are obtained for $r=0.51R$. 

Numerical evaluations of the mode shapes at point C (Fig. \ref{fig2}c) and D (Fig. \ref{fig2}e), shown in Figs.~\ref{fig2}d,f respectively for the two interfaces, confirm the localized nature of the modes and reveal their distinct distribution of the displacement magnitude and phase along the interface (see the zoomed-in plots). The magnitude displacement and phase plots suggest the possibility of selective modal excitation by applying an input at the locations shown in the figures, which highlight the spatial separation of the maximum amplitude points for the two modes. For example, preferential excitation of mode $\Psi^+$ ($\Psi^-$) could be achieved by injecting a perturbation at locations where the motion of the interface unit cells is high, and where the displacement for the other mode $\Psi^-$ ($\Psi^+$) is small (see zoomed-in views in Figs. \ref{fig2}d,f).

To confirm the splitting of the topologically protected helical waves, we designed and fabricated a waveguide made of 35 (in the $\bm{a}_1$ direction) $\times$ 25 (in the $\bm{a}_2$ direction) unit cells hosting the three different domains $H^+$ (green boundary), $H^-$ (blue boundary) and $V$ (red boundary), as shown in Fig.~\ref{fig3}a. These domains are separated by three interfaces: $\mathcal{I}(H^+,H^-)$, $\mathcal{I}(H^+,V^R)$ and $\mathcal{I}(H^-,V^r)$. Such an arrangement is chosen to illustrate the ability of the waveguide to split the two helical waves ($\Phi^+$ and $\Phi^-$) at the $y$-shaped junction. The plate is made of aluminum and the unit cell lattice parameter is $a=20.5$ mm.

First, numerical simulations are conducted to evaluate the distinct propagation patterns followed by the edge modes along the interfaces, depending on the selective mode excitation. A Finite Element (FE) model for the finite plate shown in Fig.~\ref{fig3}a is implemented in ABAQUS. Calculations are conducted in the frequency domain. Elastic waves are excited by imposing an out-of-plane harmonic excitation at half of the $\mathcal{I}(H^+,H^-)$ interface (white dot in Figs. \ref{fig3}c and \ref{fig3}d) according to the $\Psi^+$ and $\Psi^-$ configurations presented in Figs. \ref{fig2}d and \ref{fig2}f. The frequency content of the excitation is set to 98 kHz, so to prevent the excitation of bulk modes. The resulting distribution of the von Mises stress fields, reported in Figs. \ref{fig3}c,d, clearly show that when the wave reaches the $y$-shaped junction it follows either the $\mathcal{I}(H^+,V^R)$ or $\mathcal{I}(H^-,V^r)$ interface depending on the initial type of input. In both cases, weak penetration inside the bulk region is observed. Refer to SM~\cite{SM} for additional transient dynamic simulations.

The splitting of these modes is then demonstrated experimentally testing the plate (Fig.~\ref{fig3}a) by means of a Scanning laser Doppler Vibrometer (SLDV). The SLDV measures the out-of-plane velocity component of the motion of the plate surface produced by a surface bonded piezoelectric transducer, measuring $12$ mm in diameter. The excitation is applied along the $\mathcal{I}(H^+,H^-)$ interface at the location denoted by the yellow dot in Fig.~\ref{fig3}b, and consists of a $51$-cycle sine burst modulated by a Hanning window. The center frequency is $98$ kHz which falls inside the bulk bandgap and excites both $\Phi^+$ and $\Phi^-$ waves~\cite{MiniaciPRX2018}. First, one dimensional (1D) line scans of a spatial step of $0.2$ mm are conducted along the interfaces $\mathcal{I}(H^+,V^R)$ and $\mathcal{I}(H^-,V^r)$ (the locations of line scan measurements are shown as dotted black lines in Fig.~\ref{fig3}b). A temporal window of $800$ $\mu$s is applied to the recorded signals to eliminate reflections from the plate edges. Next, the recorded signals are represented in the frequency/wavenumber domain by performing a temporal/spatial Fourier transform (2D-FT), whose magnitude is superimposed in Figs.~\ref{fig4}a to the numerical dispersion predictions (white square dots) for the $\mathcal{I}(H^+,V^R)$ and $\mathcal{I}(H^-,V^r)$ finite strips. The 2D-FT magnitude colormaps clearly confirm the numerical edge state predictions along the two interfaces, and show how the two energy spots are associated to different modes in the dispersion diagrams, confirming the wave splitting.

To fully unveil the distinct nature of the $\Psi^+$ and $\Psi^-$ modes, two fine scans are conducted over the two $2 \times 2$ unit cell areas shown in red in Fig. \ref{fig3}b. The velocity distributions at specific representative time instants $t = 842$ $\mu$s and $t = 980$ $\mu$s are shown in Fig. \ref{fig4}b, where $1$ V in the colorbar corresponds to a velocity of $20$ mm/s. The two modes feature opposite spins (clockwise/anti-clockwise) of the velocity field across the interface, which is highlighted by the black arrows drawn on the basis of the phase evolution of the measured wavefield. These representations provide further evidence of mode splitting occurring at the $y$ junction. Clear visualization of the opposite spins of the two modes along the two interfaces is obtained from the measurements time animations provided in the SM~\cite{SM}.

Finally, the 2D wavefield recorded over the region highlighted by the blue dots, and labeled as ``2D scan region" in Fig. \ref{fig3}b, illustrates the $\mathcal{I}(H^+,H^-)$ interface bounded propagation along with the splitting occurring at the $y$-junction (Fig.~\ref{fig4}f). Specifically, the measured out-of-plane velocity distribution at an instant of time after the wave splitting, i.e. for $t = 1120 $ $\mu$s from the excitation, is reported. The wavefield at the considered instant of time is then represented in the wavenumber domain by performing a spatial/spatial 2D-FT, which effectively illustrates the modal content of the wavefield in the reciprocal space $k_x,k_y$. The results of this analysis shown in the 2D-FT amplitude contours of Fig.~\ref{fig4}c, illustrates the presence of 2 pairs of diffraction peaks, each corresponding to two distinct modes that co-exist at the excitation frequency, and are characterized by two distinct wavenumbers, $k_1=60$ rad/m and $k_2=80$ rad/m. These wavenumbers are highlighted by the red and black circles of different radii in Fig.~\ref{fig4}c, and correspond to the two wavenumber values associated with modes $\Psi^+$ and $\Psi^-$, respectively. The contribution of the two modes to the wavefield of Fig.~\ref{fig4}f can be effectively separated through wavenumber filtering in reciprocal space~\cite{RuzzeneSMS2007}. To this end, the 2D-FT for the wavefield is masked by 2D Gaussian windows (see Methods) centered at $k_1$ and $k_2$, whose application leads to the filtered 2D-FT in Figs.~\ref{fig4}d,e showing the two separated modes. Inverse 2D-FT transformation in physical space provides the decoupled contributions to the wavefield shown in Figs.~\ref{fig4}g,h. From these figures it clearly emerges that when the two rightward-propagating helical modes $\Phi^+$ and $\Phi^-$ (Fig. \ref{fig2}a) reach the $y$-shaped junction, they split and respectively follow the $\mathcal{I}(H^+,V^R)$ and $\mathcal{I}(H^-,V^r)$ interface as $\Psi^+$ and $\Psi^-$ modes on the basis of their polarization.

In conclusion, for the first time we proposed and experimentally tested a platform that supports multiple classes of topological modes. In the proposed configuration, implemented on a patterned plate, topologically non-trivial gaps are obtained by creating interfaces between material phases that selectively break spatial inversion symmetries. Through engineering of the nontrivial gaps, the considered system is capable of splitting purely topological protected helical edge waves into heterogeneous helical-valley modes on the basis of the initial polarization. The results presented herein, both numerical and experimental, provide fundamental insights in the behavior of topologically protected edge modes in elastic systems, and suggest new avenues for topologically protected wave transmission that may be extended to other physical domains, such as acoustics, and photonics. The findings of this study have direct implications for applications where selective waveguiding, or the isolation and control of vibrations are ultimate goals, as in civil, mechanical and aerospace engineering structures. Also, the wave mode selective capabilities of the considered interfaces and $y$-junction may be of significance for the transmission of information through elastic or acoustic waves as, for example, in the case of surface acoustic wave (SAW) devices.

\subsection*{Acknowledgments}
M.M. has received funding from the European Union's Horizon 2020 research and innovation programme under the Marie Sk{\l}odowska-Curie grant agreement N. 754364.
R.K.P. and M.R. acknowledge the support of the EFRI Award 1741685 from the National Science Foundation.

%


\subsection*{Methods}

\small{\textbf{Simulations}. Dispersion diagrams and mode shapes presented in Figs. \ref{fig1}d-f and Fig. \ref{fig2} are computed using Bloch-Floquet theory in full 3D FEM simulations carried out via the Finite Element solver COMSOL Multiphysics. Full 3D models are implemented to capture all possible wave modes supported by the plate structure. A linear elastic constitutive law is adopted and the following mechanical parameters used for the plate material (aluminum):
density $\rho = 2700$ kg/m$^3$, Young modulus $E = 70$ GPa, and Poisson ratio $\nu = 0.33$. The elastic domain is meshed by means of 8-node hexaedral elements of maximum size $L_{FE} = 0.5$ mm, which is found to provide accurate eigen solutions up to the frequency of interest \cite{miniaci2015complete}.\\
The band structures shown in Fig. \ref{fig1}d-f are obtained assuming periodic conditions along the lattice vectors $\textbf{a}_1$ and $\textbf{a}_2$. Dispersion diagrams shown in Figs. \ref{fig2} are computed instead considering a $20 \times 1$ $\textbf{a}_1$-periodic strip. The resulting eigenvalue problem $(\mathbf{K}-\omega^2 \mathbf{M})\mathbf{u} = \mathbf{0}$ is solved by varying the non-dimensional wavevector $\textbf{k}$ along the boundaries of the irreducible Brillouin zone $ \left[ \Gamma, M, K \right] $ for dispersion diagrams in Fig. \ref{fig1}d-f and within $ \left[ -\pi, \pi \right]$ for band structures presented in Figs. \ref{fig2}.\\
The distribution of the von Mises stress fields reported in Figs. \ref{fig3}c,d are conducted in the frequency domain via the Finite Element solver ABAQUS. Free boundary conditions are applied at the edges of waveguide.

\vspace{0.5cm}
\textbf{Experimental measurements and data processing}. The plate, consisting of 35 (in the $\bm{a}_1$ direction) $\times$ 25 (in the $\bm{a}_2$ direction) unit cells, is fabricated through a two-step machining process. First, the triangular holes are obtained through water-jet cutting. Circular blind and through holes are then obtained via a computer assisted drilling process.
The specimen is made of aluminum 6082 T6, with the following nominal properties: density $\rho = 2700$ kg/m$^3$, Young modulus $E = 70$ GPa, and Poisson ratio $\nu = 0.33$. The plate dimensions and key geometrical parameters are as follows: $a = 20.5$ mm, $H = 5.9 $ mm, $h = 0.5$ mm, $R = 1.75$ and $r = 0.8$ mm. Elastic waves are excited through a piezoelectric disk ($12$ mm diameter) bonded to the top surface of the plate at location shown in Fig.~\ref{fig3}b. Ultrasonic pulses consisting of $51$ sine cycles modulated by a Hanning window of central frequency of $98$ kHz are used as the excitation signals.\\
The experimental wavefields shown in Figs. \ref{fig4}b and \ref{fig4}f are recorded by a scanning laser Doppler vibrometer (SLDV) that measures the out-of-plane velocity of points belonging to a predefined grid over the structure. The spatial resolution of the grid is approximately $0.2$ mm for the 2D local scan represented by the red dotted area in Fig.~\ref{fig3}b and $0.6$ mm for the 2D local scan in the blue dotted area also in Fig.~\ref{fig3}b.

The frequency/wavenumber representation of the edge modes presented in Figs. \ref{fig4}a are obtained by performing a temporal and spatial Fourier transform (2D-FT) of the signals detected along the 1D-scan lines reported as black dotted lines in Fig.~\ref{fig3}b. The wavenumber content of the 2D scan wavefield shown in Fig. \ref{fig4}f-h are obtained by performing spatial 2D-FT of the acquired data interpolated over a regular square grid. Filtering in the wavenumber domain~\cite{RuzzeneSMS2007} for modal separation relies on the application of 2D Gaussian windowing functions centered at wavenumber $k_i$, which ca be expressed as follows: 
\[H_i(k_x,k_y) = e^{\frac{-(k-k_i)^2}{2\sigma^2}}\]
where $i=1,2$, with $k_1 = 60$ [rad/m], $k_2 = 80$ [rad/m], $k = \sqrt{k_x^2 + k_y^2}$, $\sigma^2 = 50$. The wavefields corresponding to the separated modal contributions shown in Figs.~\ref{fig4}g,h are obtained through an inverse 2D-FT of the filtered wavenumber representations shown in Figs.~\ref{fig4}d,e.

\subsection*{Data availability}

The data that support the plots within this paper and other findings of this study are available from the corresponding author upon request.

\subsection*{Contributions}
All  authors contributed extensively to the work presented in this paper.

\subsection*{Competing interests}
The authors declare no competing financial interests.

\clearpage

\begin{figure}
\centering
\begin{minipage}[]{1\linewidth}
{\includegraphics[trim={0in 0in 0in 0in}, width=1\textwidth]{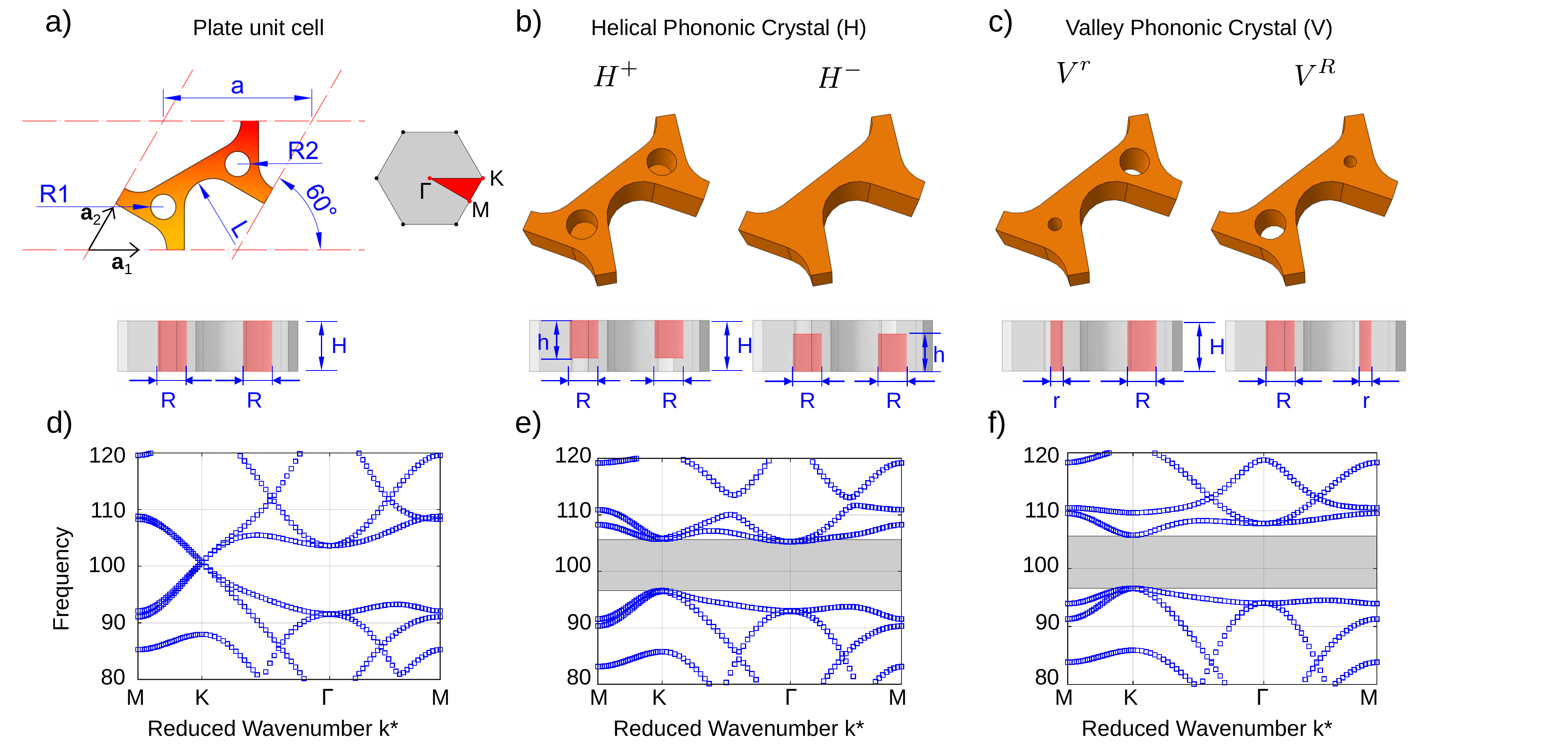}} 
\end{minipage}
\caption{\textbf{Design of the unit cells leading to distinct topological phases and their dispersion properties}.
\textbf{a}, In-plane and cross-sectional view of the unit cell with through holes. The holes have equal diameter $R = 0.0875 a$, where $a = 20.5$  mm is the magnitude of the lattice vectors ($a=\bm{a}_1=\bm{a}_2$) and $H = 5.9$ mm denotes the plate thickness. The inset shows the irreducible Brillouin zone and the high symmetry points $\Gamma$, $K$ and $M$.
\textbf{b}, Perspective and cross-sectional view of the unit cells ($H^+$ and $H^-$) emulating spin orbital coupling in Quantum spin Hall effect with $\sigma_h$ broken symmetry (blind holes).
\textbf{c}, Perspective and cross-sectional views of the unit cells ($V^r$ and $V^R$) emulating the Quantum valley Hall effect with $\sigma_v$ broken symmetry (through holes of radii $r \neq R$).
\textbf{d}-\textbf{f}, Calculated phononic band structure for the plate with through holes, and for the plates composed of $H^+$ ($H^-$) and $V^r$ ($V^R$) unit cells, respectively. The plate with through holes is characterized by a double degenerate Dirac point visible in (d), while the cases of $H^+$ ($H^-$) (e) and $V^r$ ($V^R$) (f) feature a complete bandgap (light gray rectangle) centered at approximately $102$ kHz. The widths and center frequency of the bandgaps are matched by selecting the partial depth of the blind holes $h$ in the $H^+$ ($H^-$) configuration ($h=0.91 H$ for the band diagram shown), and the radii $r$ and $R$ of the through holes in the $V^r$ ($V^R$) case ($r=0.51R$ for the diagrams shown). Refer to the Methods section \textit{Simulations} for details on band structure calculations.}
\label{fig1}
\end{figure}

\begin{figure}
\centering
\begin{minipage}[]{1\linewidth}
{\includegraphics[trim={0.125in 0cm 4.5in 0.25in}, width=0.8\textwidth]{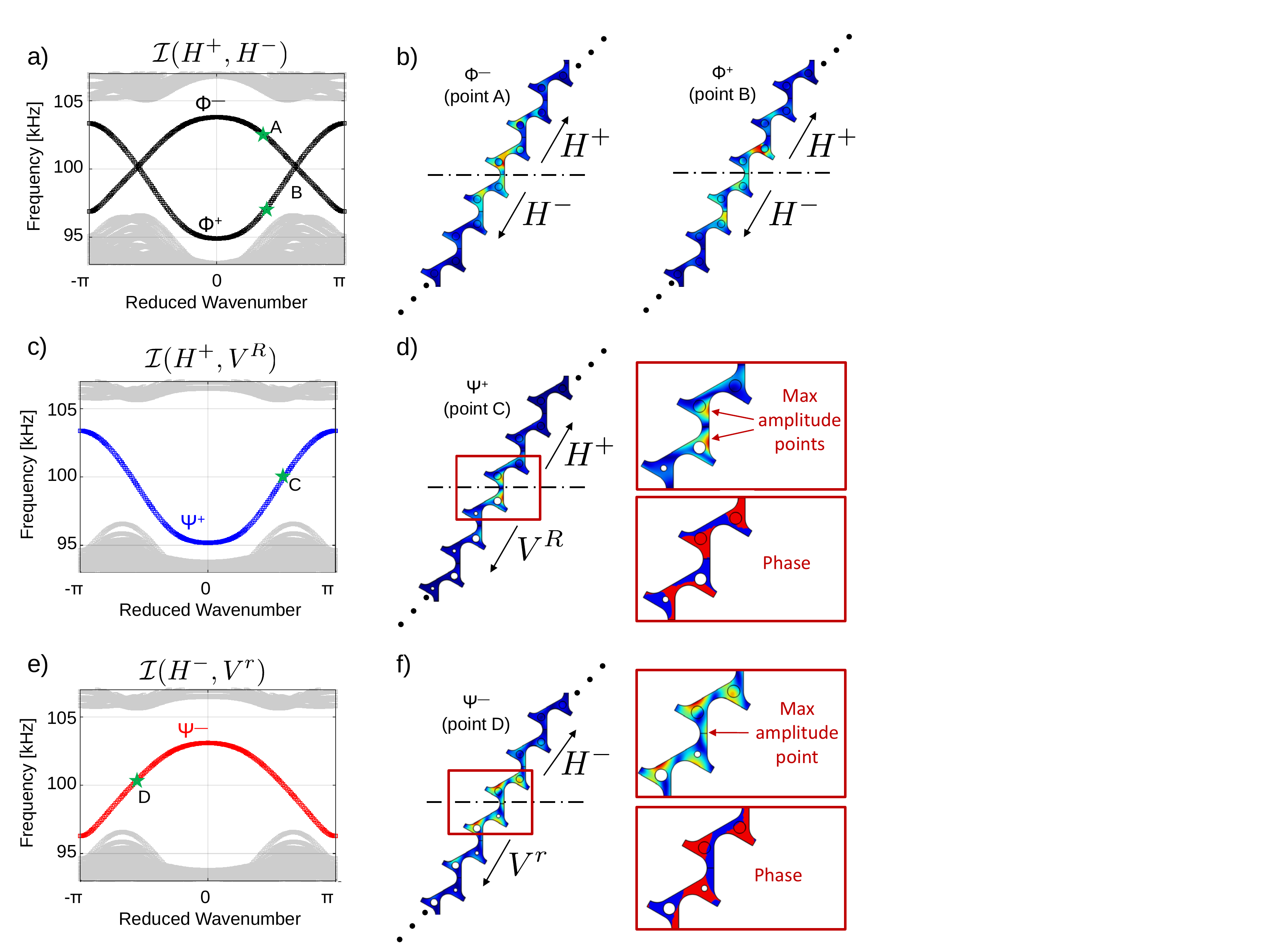}}
\end{minipage}
\caption{\textbf{Non-trivial interfaces: band structure and edge states}
\textbf{a,c,e}, Dispersion diagrams for the non-trivial waveguides defined by $\mathcal{I}(H^+,H^-)$, $\mathcal{I}(H^+,V^R)$, and $\mathcal{I}(H^-,V^r)$ interfaces. The band structures are computed considering a $20 \times 1$ $\bm{a}_1$-periodic strip ($10$ unit cells on each side of the domain wall). The bulk modes are reported as gray dots while the interface modes in black, blue and red dotted lines, respectively. The edge modes are denoted by the index $+$ ($-$) according to the positive (negative) group velocity relative to the propagation direction.
\textbf{b,d,f}, Corresponding eigenvectors (colors represent magnitudes of the absolute normalized displacement, varying from zero (blue) to maximum (red)) show mode localization at the interface (the deformation for only $6$ cells is reported for the clarity of representation).
\textbf{c,d} Close-ups for the $\Psi^+$ and $\Psi^-$ modes highlight the different displacement distribution and phases (positive in blue and negative in red) of the modes at the interface, which suggests the possibility of selective mode excitation. The preferential mode excitation for $\Psi^+$ ($\Psi^-$) can be achieved by applying an excitation at the maximum amplitude point highlighted in the insets.}
\label{fig2}
\end{figure}

\begin{figure}
\centering
\begin{minipage}[]{1\linewidth}
{\includegraphics[trim={0in 4.375in 0in 1.875in}, width=1\textwidth]{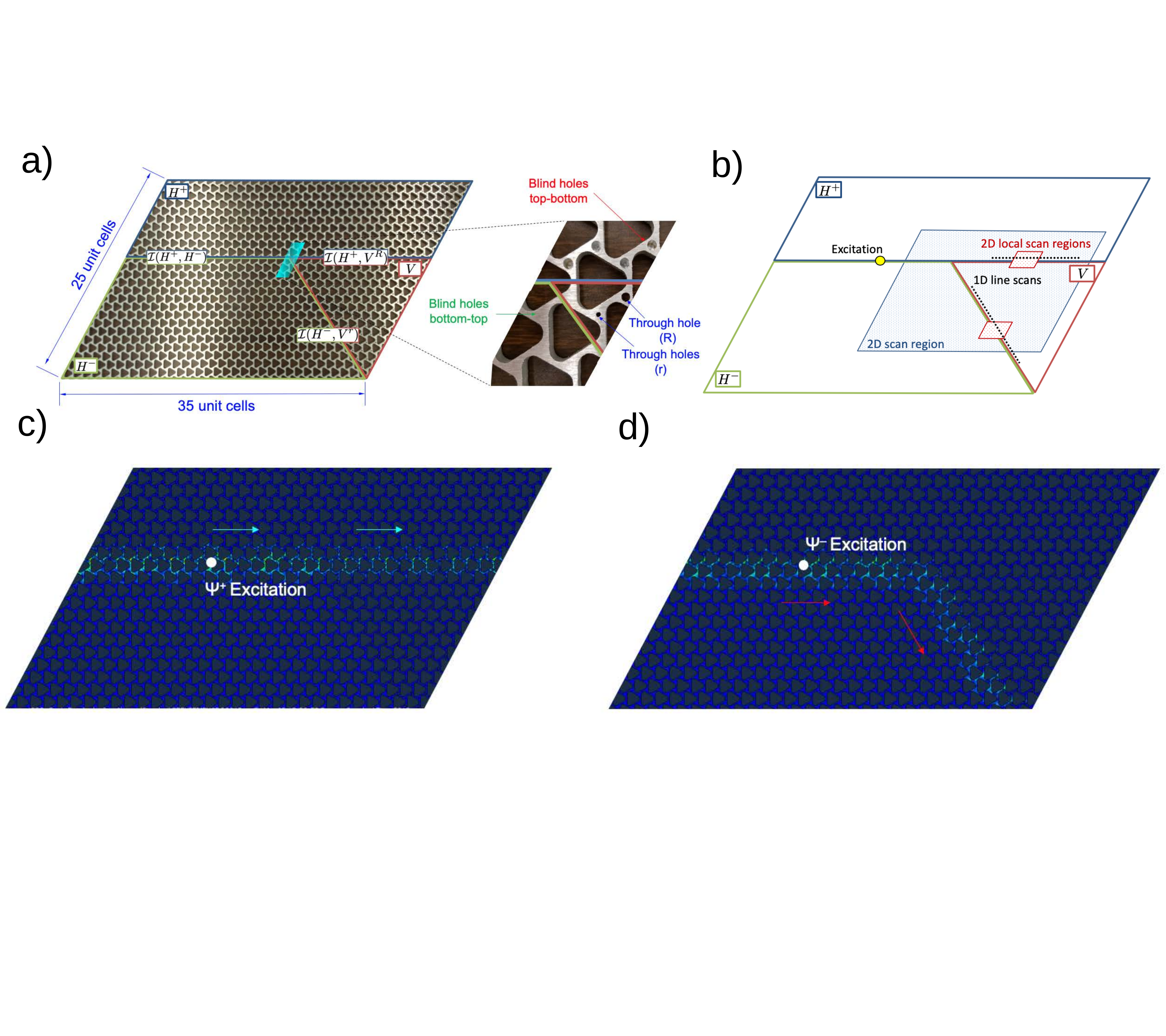}}
\end{minipage}
\caption{\textbf{Configuration of the finite structure and numerical simulations showing selective mode waveguiding.}
\textbf{a}, Schematic representation and experimental implementation of the non-trivial waveguide hosting $H^+$ (in red), $H^-$ (in green) and $V$ (in blue) phases giving rise to $3$ interfaces:  $\mathcal{I}(H^+,H^-)$ (blue-green), $\mathcal{I}(H^+,V^R)$ (blue-red), and $\mathcal{I}(H^-,V^r)$ (red-green).
\textbf{b}, Waveguide schematic showing the locations of the excitation and 1D and 2D scan points/regions considered in the experiments (1D scans are the black dotted lines, local 2D scans are the red dotted areas, while large 2D scan of the $y$-junction region is the blue dotted area). 
\textbf{c,d}, Numerical distribution of the von Mises stress field resulting from harmonic excitation at 98 kHz, i.e. within the bulk gap. The excitation is applied at the location shown by the white dot as an out-of-plane displacement distribution. Specific displacement distribution of the surface stress is applied according to the modal content of the modes shown in Fig.~\ref{fig2}d,f in order to selectively induce mode $\Psi^+$ (c) and $\Psi^-$ (d), respectively. The calculations clearly illustrate the possibility to preferentially excite one of the two modes and to remotely select the  interface along which the wave propagates. Colors indicate the von Mises stress magnitude, ranging from zero displacement (blue) to maximum displacement (red). Refer to SM \cite{SM} for additional transient dynamic simulations.}
\label{fig3}
\end{figure}

\begin{figure}
\centering
\begin{minipage}[]{1\linewidth}
{\includegraphics[trim={0.25in 3.125in 0.125in 5.5in}, width=0.95\textwidth]{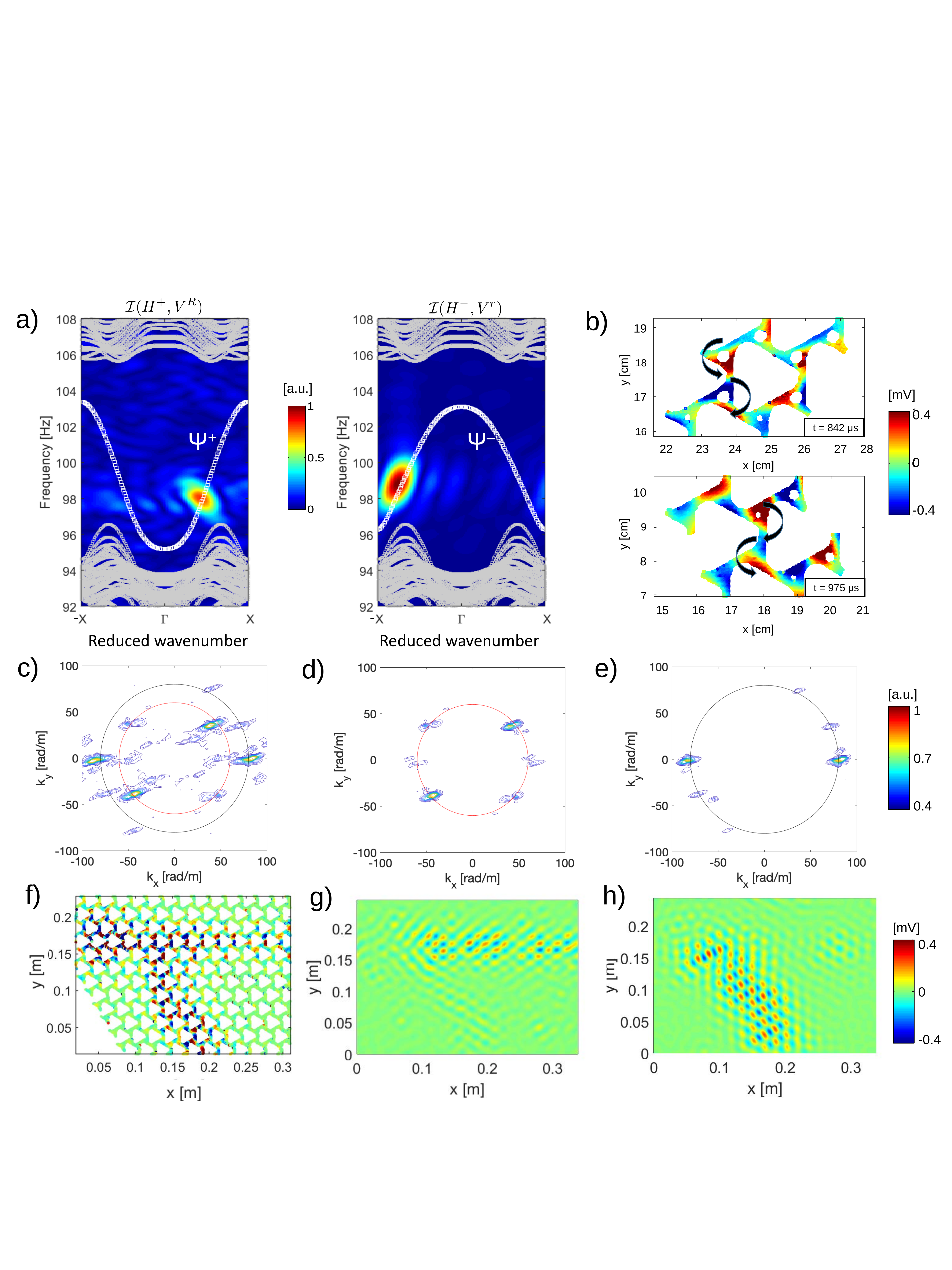}}
\end{minipage}
\caption{\textbf{Experimental observation.}
\textbf{a}, Spatio/temporal 2D-FT of 1D line scans along interfaces $\mathcal{I}(H^+,V^R)$ and $\mathcal{I}(H^-,V^r)$. Colormaps show normalized 2D-FT amplitudes, superimposed to the numerically predicted band structure (white square dots) highlighting the $\Psi^+$ (left panel) and $\Psi^-$ (right panel) modes.
\textbf{b}, Measured wavefield for the $\Psi^+$ and $\Psi^-$ propagating edge modes displaying opposite spins profiles, as highlighted by the superimposed black arrows. The measurements correspond to the area denoted as ``2D local scan regions" in Fig. \ref{fig2}b. Time animations of the measured wavefields are provided in the SM~\cite{SM}.
\textbf{c}, Spatial 2D-FT for a representative snapshot ($t = 1120 \mu$) of the wavefield recorded over the ``2D scan region" in Fig.~\ref{fig3}b and shown in Fig.~\ref{fig4}f. The 2D-FT highlights the presence of a pair of diffraction peaks defined by the concentration at contour levels in the reciprocal space $k_x,k_y$, which are associated with wavenumbers $k_1=60$ rad/m (red circle) and $k_2=80$ (black circle) rad/m, corresponding to the distinct modes of propagation $\Psi^+$ and $\Psi^-$.
\textbf{d,e} Filtered 2D-FTs with isolated modes and \textbf{g,h} corresponding propagation in physical space showing the decoupled wavefields and distinction propagation paths for the two separated modes.}
\label{fig4}
\end{figure}

\end{document}